\documentclass[aps,pre,superscriptaddress,amsmath,amssymb,reprint]{revtex4-2}

\usepackage{graphicx}
\usepackage{txfonts}
\usepackage{mathtools}
\usepackage[hypertexnames=false]{hyperref}
\usepackage{braket}
\usepackage{bm}
\usepackage{xcolor}

\newcommand{\aln}[1]{\begin{align}#1\end{align}}
\newcommand{\nn}{\nonumber\\}

\begin{document}
\title{Loop expansion in polymer field theory: application to phase separation}

\author{Kiyoharu Kawana}
\email{kkiyoharu@kias.re.kr}
\affiliation{School of Physics, Korea Institute for Advanced Study, Seoul 02455, Korea}

\author{Kyosuke Adachi}
\email{kyosuke.adachi@riken.jp}
\affiliation{RIKEN Center for Interdisciplinary Theoretical and Mathematical Sciences, 2-1 Hirosawa, Wako 351-0198, Japan}
\affiliation{RIKEN Pioneering Research Institute, 2-1 Hirosawa, Wako 351-0198, Japan}
\affiliation{RIKEN Center for Biosystems Dynamics Research, 2-2-3 Minatojima-minamimachi, Chuo-ku, Kobe 650-0047, Japan}

\date{\today}

\begin{abstract}
Liquid-liquid phase separation underlies phenomena ranging from protein condensate formation to the phase coexistence of synthetic polymers.
Although the random phase approximation (RPA) is widely used to predict such phase behavior, its quantitative accuracy for binodals of polymer solutions, particularly outside the high-density regime, remains incompletely characterized.
Here, we develop a field-theoretic loop expansion in homopolymer systems by identifying the inverse polymer density $\rho^{-1}$ as the Planck constant $\hbar$ in quantum field theory.
We calculate the leading-order and next-to-leading-order corrections to the RPA free energy, denoted as RPA+ and RPA++, respectively.
Testing the binodal predicted by the RPA+ against molecular dynamics simulations of bead-spring chains with Gaussian pair interactions, we find that the RPA+ qualitatively improves the dilute-phase coexistence density over the RPA, while the critical point error remains comparable to that of the RPA.
Our results establish the loop expansion as a systematic route for refining the RPA-based binodal predictions for polymer phase separation.
\end{abstract}
\maketitle

\section{Introduction}\label{Sec:intro}

Liquid-liquid phase separation of disordered protein polymers has been recognized as a fundamental principle of intracellular organization, underlying the formation of membraneless compartments such as nucleoli and stress granules~\cite{Brangwynne2015, Banani2017, Berry2018, Choi2020}.
Quantitative prediction of phase behavior observed in experiments~\cite{Martin2020, Bremer2022}, in particular the binodal curve for phase separation, is a central challenge for coarse-grained polymer models of disordered proteins~\cite{Feito2025, Rizuan2026}.
Apart from molecular dynamics simulations of such models, field-theoretic approaches have been employed for numerical sampling of equilibrium states~\cite{Fredrickson2002, Fredrickson2023, McCarty2019}, as well as for analytical calculation of binodals with low computational cost~\cite{Fredrickson2006, Lin2016, Ellis2026}.
Even though molecular dynamics and field-theoretic simulations are more accurate, such analytical theories remain valuable; they allow rapid scanning over large parameter spaces such as chain length and interaction parameters and provide closed-form thermodynamic quantities that clarify which parameters control phase behavior.

The random phase approximation (RPA), obtained by the Gaussian (i.e., $1$-loop) approximation for fluctuations around the homogeneous mean-field saddle point within the field-theoretic approach~\cite{Edwards1966,Fredrickson2006}, has been a powerful analytical tool to predict polymer phase separation.
The RPA has been extensively applied to polyelectrolyte solutions~\cite{Borue1988, Mahdi2000, Popov2007, Delaney2017} and solutions of polyampholytes or charged disordered proteins~\cite{Lin2016, Lin2017, McCarty2019}.
Refinements of the RPA, such as renormalization of the Kuhn length~\cite{Lin2020}, have also been proposed to improve the prediction accuracy.
More generally in polymer physics, improvements of RPA have been broadly discussed, e.g., the Hartree-level~\cite{Muthukumar1982} and renormalization-group~\cite{Ohta1983} approaches for polymer solutions in good solvents, as well as renormalization of divergent contributions in the RPA for polymer blends and melts~\cite{Grzywacz2007}.

Despite such developments, the RPA-based methods for predicting the phase separation of polymer solutions have been applied mainly to systems with long-range electrostatic interactions, while short-range repulsive and attractive interactions are often incorporated at the mean-field level~\cite{Lin2016, Wessen2022}.
Though the RPA should be accurate at high polymer densities even for systems with short-range interactions~\cite{Doi1986, Fredrickson2006}, it is unclear how well it can predict the binodal curve, which is determined by the free energy in a wide range of densities.
To develop an analytical framework applicable to both long-range and short-range interactions, and to systematically improve the accuracy of the RPA, it is natural to go beyond the Gaussian approximation and consider higher-order corrections, as previously discussed in polymer theory~\cite{Muthukumar1982, Ohta1983}.

In quantum field theory, the loop expansion corresponds to an expansion in powers of the Planck constant $\hbar$, which clearly indicates that the limit $\hbar\rightarrow 0$ corresponds to the classical (or saddle-point) limit.
This naturally raises the question: What is the expansion parameter in the field-theoretic approach in polymer systems?
In the following, we explicitly show that the inverse of the polymer density $\rho^{-1}$ plays the role of $\hbar$.
This correspondence implies that the RPA is increasingly accurate in the high-density limit $\rho\rightarrow \infty$, as anticipated from the previous studies~\cite{Doi1986, Fredrickson2006}.

Once we understand the correspondence $\hbar~\leftrightarrow~\rho^{-1}$, it is clear how the RPA can be systematically improved: the loop expansion leads to the higher-order corrections to thermodynamic quantities in powers of $\rho^{-1}$.
In this paper, we identify the diagrams corresponding to the leading-order~(LO) and next-to-leading-order (NLO) corrections, and evaluate these diagrams explicitly.
For comparison with the conventional RPA, we refer to the LO and NLO analyses as RPA+ and RPA++, respectively.
We then test the binodals predicted by the RPA+ against molecular dynamics (MD) simulations of homopolymer systems with short-range interactions.
We find that the RPA+ qualitatively improves the RPA prediction of the dilute-phase density, bringing it to the same order of magnitude as observed in simulations.
On the other hand, the prediction near the critical point is not improved by the RPA+, indicating the need for incorporating alternative methods, such as the renormalization-group method, that can effectively resum higher-loop corrections.
Taken together, our results establish the loop expansion as a systematic route to refine RPA-based binodal predictions for polymer phase separation.

We note that a systematic diagrammatic expansion of the polymer field theory was formulated by Morse~\cite{Morse2006}, who pointed out through power counting that each additional loop carries one inverse power of a concentration-type parameter.
Compared with this, the novelty of the present work lies in (i) establishing the correspondence to quantum field theory explicitly, with $\rho^{-1}$ playing the role of $\hbar$, and (ii) explicitly evaluating the loop corrections and thereby the binodal of a polymer solution, tested directly against MD simulations.

The organization of this paper is as follows.
In Sec.~\ref{sec_single}, we summarize the basic properties of a single polymer and introduce the density correlation functions that serve as the building blocks of the polymer field theory.
In Sec.~\ref{sec_equilibrium}, we formulate the polymer field theory for an equilibrium system of homopolymers and review the RPA.
In Sec.~\ref{sec_beyond}, we develop the loop expansion in the polymer field theory and evaluate the LO (RPA+) and NLO (RPA++) corrections.
In Sec.~\ref{sec_comparison}, we compare the RPA and RPA+ binodal predictions with MD simulations of a homopolymer solution with short-range Gaussian interactions.
Section~\ref{sec_conclusion} summarizes our findings and discusses future directions.

\section{Single polymer system}
\label{sec_single}

In the field-theoretic approach for equilibrium polymers, the information of monomer correlation functions in a single polymer~\cite{Yamakawa1971} is important.
Let us summarize the fundamental results in a single polymer system.
In the following, we denote a point in $D$-dimensional space simply by $x=(x^1,\cdots,x^D)$.

The single-polymer partition function is defined by
\aln{
Q[\phi]\coloneqq \frac{\left(\prod_{i=1}^N\int d^Dx_i\right)\exp\left(-\beta \sum_{l=1}^{N-1}u(r_l^{})+\sum_{l=1}^N\phi(x_l^{})\right)}{\left(\prod_{i=1}^N\int d^Dx_i\right)\exp\left(-\beta \sum_{l=1}^{N-1}u(r_l^{})\right)}~,
\label{single polymer}
}
where $\beta=(k_B T)^{-1}$ is the inverse temperature with $k_B^{}$ the Boltzmann constant, $N$ is the total number of monomers (i.e., the degree of polymerization), $r_l^{}=x_{l+1}^{}-x_l^{}$, $u(x)$ is a general interaction between adjacent monomers, $-\phi(x)$ is an external potential, and we do not consider non-nearest-neighbor interactions.
Note that $Q[\phi]$ is normalized by the field-free single-polymer partition function such that $Q[0]=1$.
Note also that the sign in front of $\phi$ in the exponent is opposite to the convention of Ref.~\cite{Fredrickson2006}; accordingly, $Q[-i\phi]$ appearing below corresponds to $Q[i\phi]$ in that convention.
Throughout this paper, we choose the Gaussian chain
\aln{\beta u(x)=\frac{Dx^2}{2b^2}~,
}
where $b$ is a bond length, as often used in the literature~\cite{Doi1986, Fredrickson2006}.

By introducing the monomer density
\aln{
\label{monomer density}
\hat{\rho}_S^{}(x)\coloneqq \sum_{i=1}^{N}\delta^{(D)}(x-x_i^{})=\int \frac{d^Dk}{(2\pi)^D}\sum_{i=1}^N e^{ik\cdot (x-x_i^{})}~,
}
we can also express the external potential term as
\aln{\sum_{i=1}^N\phi(x_i^{})=\int d^Dx~\hat{\rho}_S^{}(x)\phi(x)~.
}
Here, the subscript $S$ means ``single" polymer.
Then, we can introduce the $n$-point density correlation functions as
\aln{
F^{(n)}_{\rm free}(X_1^{},\cdots,X_n^{})&\coloneqq \frac{1}{Q[\phi]}\frac{\delta^n Q[\phi]}{\delta \phi(X_1^{})\cdots \delta \phi(X_n^{})}\bigg|_{\phi=0}
\\
&=\langle \hat{\rho}_{S}^{}(X_1^{})\hat{\rho}_{S}^{}(X_2^{})\cdots \hat{\rho}_{S}^{}(X_n^{})\rangle_{\rm free}^{}~,
}
where $\langle \cdots \rangle_{\rm free}^{}$ means the expectation value in the free single-polymer system, namely, $\phi=0$.
By using Eq.~(\ref{monomer density}), this becomes
\aln{
&=\sum_{i_1^{}=1}^N\cdots \sum_{i_n^{}=1}^{N}\left(\prod_{j=1}^n\int \frac{d^Dk_j^{}}{(2\pi)^D}\right)
\left\langle e^{i\sum_{j=1}^nk_j^{}\cdot (X_j^{}-x_{i_j^{}}^{})}\right\rangle_{\rm free}^{}~.
}
The corresponding Fourier mode is
\aln{
\tilde{F}_{\rm free}^{(n)}(k_1^{},\cdots,k_n^{})&\coloneqq \left(\prod_{j=1}^n \int d^DX_j^{}e^{-ik_j^{}\cdot X_j^{}} \right)F^{(n)}_{\rm free}(X_1^{},\cdots,X_n^{})
\nn
&=\sum_{i_1^{}=1}^N\cdots \sum_{i_n^{}=1}^{N}\left\langle e^{-i\sum_{j=1}^nk_j^{}\cdot x_{i_j^{}}^{}}\right\rangle_{\rm free}^{}~,
\label{n-point Fourier mode}
}
where we have used $\delta^{(D)}(k)=\int \frac{d^Dx}{(2\pi)^D}e^{ik\cdot x}$.
After some calculation~(Appendix~\ref{app:monomer correlation}), we obtain the following expression:
\aln{
\tilde{F}_{\rm free}^{(n)}(k_1^{},\cdots,k_n^{})&=(2\pi)^D\delta^{(D)}\left(\sum_{j=1}^{n}k_j^{}\right)\frac{1}{V}
\nn
\times\sum_{i_1^{}=0}^{N-1}&\cdots \sum_{i_n^{}=0}^{N-1} \exp\left(-\frac{b^2}{2D}\sum_{j=1}^{n}\sum_{j'=1}^{n}k_j^{}\cdot k_{j'}^{}\mathrm{min}(i_j^{},i_{j'}^{})
\right)~,
\label{Fourier mode correlations}
}
where we have replaced $i_j^{}-1\rightarrow i_j^{}$, and $V$ is the space volume.
Here and below, we borrow the terminology of quantum field theory and refer to the wave vector $k_j^{}$, which is conjugate to the real-space coordinate $X_j^{}$ through the Fourier transform, as ``momentum''.
The prefactor $(2\pi)^D\delta^{(D)}\left(\sum_{j=1}^{n}k_j^{}\right)$ in Eq.~(\ref{Fourier mode correlations}) arises from the translational invariance of the single-polymer ensemble,
% [see Eq.~(\ref{n-point Fourier mode}) and Appendix~\ref{app:monomer correlation}]:
i.e., the Fourier-mode correlation function is nonzero only when the wave vectors carried by all the legs add up to zero.
By analogy with quantum field theory, we call this constraint ``momentum conservation''.
% and we refer to $(2\pi)^D\delta^{(D)}\left(\sum_{j=1}^{n}k_j^{}\right)$ as the momentum-conservation factor throughout this paper.No mechanical momentum is involved; the name is used only for convenience.

\begin{figure}
\begin{center}
\includegraphics[scale=0.6]{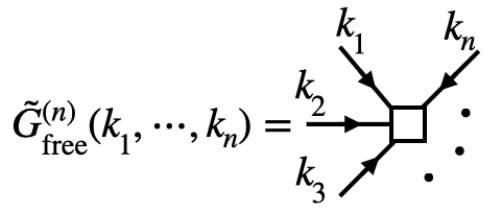}
\caption{
A graphical representation of the connected correlation function $G^{(n)}_{\rm free}$.
Here, the square implies that it corresponds to a connected correlation function, and the dots represent other unshown legs.
}
\label{fig:vertex}
\end{center}
\end{figure}

We can also introduce the connected density correlation functions and their Fourier modes by taking functional derivatives of $\log Q[\phi]$:
\aln{
G^{(n)}_{\rm free}(X_1^{},\cdots,X_n^{})&\coloneqq \frac{\delta^n \log Q[\phi]}{\delta \phi(X_1^{})\cdots \delta \phi(X_n^{})}\bigg|_{\phi=0}~,
\\
\tilde{G}_{\rm free}^{(n)}(k_1^{},\cdots,k_n^{})&\coloneqq \left(\prod_{j=1}^{n} \int d^DX_j^{}e^{-ik_j^{}\cdot X_j^{}} \right)G^{(n)}_{\rm free}(X_1^{},\cdots,X_n^{})~.
}
Factoring out the momentum-conservation factor, $(2\pi)^D\delta^{(D)}\left(\sum_{j=1}^n k_j^{}\right)/V$, we express these Fourier modes as
\aln{
M\tilde{F}_{\rm free}^{(n)}(k_1^{},\cdots,k_n^{})\equiv \rho(2\pi)^D \delta^{(D)}\left(\sum_{i=1}^n k_i^{}\right)\tilde{f}_{\rm free}^{(n)}(k_1^{},\cdots,k_n^{})~,
\label{Fourier modes of nonconnected}
\\
M\tilde{G}_{\rm free}^{(n)}(k_1^{},\cdots,k_n^{})\equiv \rho (2\pi)^D\delta^{(D)}\left(\sum_{i=1}^n k_i^{}\right)\tilde{g}_{\rm free}^{(n)}(k_1^{},\cdots,k_n^{})~,
\label{Fourier modes of connected}
}
where $M$ denotes the total number of polymers and $\rho=M/V$ is the polymer density.
In particular, one can check that the zero mode is
\aln{
\tilde{f}_{\rm free}^{(n)}(0,\cdots,0)=N^n~
\label{connected zero modes}
}
by the definition~(\ref{Fourier mode correlations}).
In this paper, we represent the connected correlation function $G^{(n)}_{\rm free}$ graphically as Fig.~\ref{fig:vertex}.
The connected correlation function $G^{(n)}_{\rm free}$, which is often called the cumulant (or cumulant moment) in the polymer literature~\cite{Fredrickson2006, Morse2006}, collects the fully connected part of the density correlations.
These connected functions serve as the natural building blocks (i.e., vertices) of the loop expansion developed in Sec.~\ref{sec_beyond}.

Since these correlation functions are defined by the same functional $Q[\phi]$, there exist rigorous relations among them~\cite{Peskin1995}, which enable us to express the connected correlation functions $\{G_{\rm free}^{(n)}\}$ by the non-connected ones $\{F_{\rm free}^{(n)}\}$.
In Appendix~\ref{app:correlation}, we summarize these relations for up to $n=4$.

\section{Equilibrium polymers}
\label{sec_equilibrium}

Let us consider the equilibrium system of homopolymers.
A Hamiltonian of $M$ polymers is given by the potential energy
\aln{\label{polymer potential}
H_M^{}(\{x_{\alpha i}^{}\})=\sum_{\alpha=1}^M\sum_{i=1}^{N-1}u(r_{\alpha i}^{})+\frac{1}{2}\sum_{(\alpha,i)\neq (\beta,j)}v(x_{\alpha i}^{}-x_{\beta j}^{})~,
}
where $\alpha,\beta,\cdots~(i,j,\cdots )$ are the indices of polymers (monomers), $r_{\alpha i}^{}\coloneqq x_{\alpha i+1}^{}-x_{\alpha i}^{}$, and $\sum_{(\alpha,i)\neq (\beta,j)}$ denotes the sum over all different monomers.
Here, we have omitted the kinetic (momentum) part of the Hamiltonian, which is quadratic in momenta and, upon integration of the Boltzmann factor over the momenta, only produces the thermal-wavelength prefactor in Eq.~(\ref{Z definition}) with $m$ the mass of a monomer and $h$ the Planck constant.
Note that the effects of the solvent are supposed to be included in the two-body potential $v(x)$ effectively.
For the following discussion, we also introduce the total monomer density function
\aln{
\hat{\rho}(x)\coloneqq \sum_{\alpha=1}^M\sum_{i=1}^N \delta^{(D)}(x-x_{\alpha i}^{})~.
\label{density operator}
}
without any subscript.

\subsection{Polymer field theory}

Below, we explain the field-theoretic approach in the equilibrium polymer system~\cite{Fredrickson2002, Fredrickson2006, Fredrickson2023}.
We assume $v(0) < \infty$ in the following.

The canonical partition function is defined by
\aln{
Z\coloneqq&\frac{(2\pi m k_B T/h^2)^{\frac{D}{2}MN}}{M!}\left(\prod_{(\alpha,i)}\int d^Dx_{\alpha i}^{}\right)\exp\left(
-\beta H_M^{}(\{x_{\alpha i}^{}\})\right)
\label{Z definition}
\\
=&\frac{z^{M}}{M!}\left(\prod_{(\alpha,i)}\int_V d^Dx_{\alpha i}^{}\right)
\nn
\times &\exp\left(
-\beta \sum_{\alpha=1}^M\sum_{l=1}^{N-1}u(r_{\alpha l}^{})-\frac{\beta}{2}\int d^Dx\int d^Dy\hat{\rho}(x)v(x-y)\hat{\rho}(y)\right)~,
}
where $z=e^{\frac{\beta}{2}Nv(0)}(2\pi m k_B T/h^2)^{\frac{D}{2}N}$.
Note that the factor $e^{\frac{\beta}{2}Nv(0)}$ originates from the subtraction of the self-interactions~\cite{Fredrickson2006, Fredrickson2023}.
Substituting the functional identity
\aln{
1&=\int {\cal D}\rho \delta(\rho(x)-\hat{\rho}(x))
\nn
&=\int {\cal D}\rho \int {\cal D}\phi~e^{i\int d^Dx(\rho(x)-\hat{\rho}(x))\phi(x)}~,
}
where the normalization constant of the functional integral is absorbed into the integration measure ${\cal D}\phi$,
we obtain
\aln{
Z&=Z_0^{}\int {\cal D}\rho\int {\cal D}\phi~e^{-S[\rho,\phi]}~,
\label{canonical partition function}
\\
S[\rho,\phi]&=\frac{\beta}{2}\int d^Dx\int d^Dy \rho(x)v(x-y)\rho(y)
\nn
&\quad-i\int d^Dx\rho(x)\phi(x)-M\log Q[-i\phi].
\label{Euclidean action}
}
Here,
\aln{
Z_0^{}=\frac{(2\pi m k_B T/h^2)^{\frac{D}{2}MN}}{M!}\left[V e^{\frac{\beta}{2}Nv(0)} \left(\frac{2\pi b^2}{D}\right)^{\frac{D(N-1)}{2}}\right]^M,
}
which corresponds to the partition function of an ideal gas of bead-spring polymers, i.e., $M$ non-interacting chains, and $Q[-i\phi]$ is the normalized single-polymer partition function~(\ref{single polymer}) with an imaginary external potential.
Here, the factor $V(2\pi b^2/D)^{\frac{D(N-1)}{2}}$ in $Z_0^{}$ is the field-free value of the configurational integral for a single chain, consistently with the normalization $Q[0]=1$.
Note that the free-energy density of ideal polymers is
\aln{
f_0^{}&\coloneqq -k_B^{}T\frac{\log Z_0^{}}{V}
\nn
&=k_B^{}T\rho\biggl[\log \left(\frac{\rho}{(2\pi mk_BT/h^2)^{\frac{D}{2}N}(2\pi b^2/D)^{\frac{D(N-1)}{2}}}\right)-1 - \frac{\beta}{2}Nv(0) \biggr]~.
\label{ideal polymer}
}
In evaluating $\log Z_0^{}$ we used Stirling's approximation $\log M!\simeq M\log M-M$ (with $\rho=M/V$), which holds for large $M$, i.e., in the thermodynamic limit assumed throughout this paper; in particular, the $-1$ inside the square bracket of Eq.~(\ref{ideal polymer}) originates from this approximation. 

We assume that the Fourier mode $\tilde{v}(k)$ of $v(x)$ is greater than zero for $^\forall k$.
Then we can perform the path-integral of $\rho(x)$ as
\aln{
\label{field theory}
Z&=Z_0^{}\int {\cal D}\phi e^{-S_\phi^{}}~,
\\
S_\phi^{}&=\frac{1}{2\beta}\int d^Dx\int d^D y\phi(x)v^{-1}(x-y)\phi(y)-M\log Q[-i\phi]~,
}
where $v^{-1}(x)$ is the inverse function of $v(x)$:
\aln{
\int d^Dyv(x-y)v^{-1}(y-z)=\delta^{(D)}(x-z)~.
}
The step from Eq.~(\ref{canonical partition function}) to Eq.~(\ref{field theory}) is a Hubbard--Stratonovich transformation: the term quadratic in $\rho$ in Eq.~(\ref{Euclidean action}) is a Gaussian integral, $\int{\cal D}\rho\,\exp\!\left(-\frac{\beta}{2}\int\!\int\rho v\rho+i\int\rho\phi\right)\propto\exp\!\left(-\frac{1}{2\beta}\int\!\int\phi v^{-1}\phi\right)$, so that integrating out $\rho$ trades the two-body interaction $v$ for the auxiliary field $\phi$ with a quadratic weight $\propto v^{-1}$. 
Equation~(\ref{field theory}) allows us to perform the saddle point approximation for $M\rightarrow \infty$ with the volume $V$ fixed~\cite{Fredrickson2006}.
The saddle point of $\phi(x)$ is determined by
\aln{
\frac{\delta S_\phi^{}}{\delta \phi(x)}=0~,
}
namely
\aln{
&\frac{1}{\beta}\int d^Dy v^{-1}(x-y)\phi(y)=
\nn
&-i\frac{M}{Q[-i\phi]}\left\langle \hat{\rho}_S^{}(x)\,\exp\left(-i\sum_{i=1}^N\phi(x_i^{})\right)\right\rangle_{\rm free}^{}~.
}
In particular, for $\phi(x)=\phi=\text{constant}$ (i.e., the homogeneous saddle point), this becomes
\aln{
\frac{\tilde{v}(0)^{-1}}{\beta}\phi=-i\frac{MN}{V}\quad \therefore \quad \phi=\phi_s^{}=-i\beta \tilde{v}(0)N\rho~,
}
where $\tilde{v}(0)=\int d^Dx v(x)$.

Introducing a fluctuation by
\aln{
\phi(x)=\phi_s^{}+\chi(x)~,
}
the partition function~(\ref{field theory}) is now written as
\aln{
Z=&Z_0^{}e^{VN^2\beta \frac{\tilde{v}(0)}{2}\rho^2+M\log Q[-i\phi_s^{}]}
\nn
\times&\int {\cal D}\chi~
\exp\bigg(-\frac{1}{2}\int d^Dx\int d^Dy\chi(x)
\Gamma^{(2)}(x-y)\chi(y)
\nn
&+\sum_{k=3}^{\infty}\frac{(-i)^k}{k!}\left(\prod_{i=1}^k\int d^Dx_i^{}\right)MG_{\rm free}^{(k)}(x_1^{},\cdots,x_k^{})\chi(x_1^{})\cdots \chi(x_k^{})\bigg)~,
\label{polymer field theory}
}
where
\aln{\label{kinetic term}
\Gamma^{(2)}(x)=k_BTv^{-1}(x)+MG^{(2)}_{\rm free}(x)~
}
corresponds to the kinetic term for $\chi(x)$.
Here, by ``kinetic term'' we mean the quadratic (Gaussian) part of the exponent in Eq.~(\ref{polymer field theory}), regarded as a functional of $\chi$.
As in quantum field theory, the inverse of this quadratic kernel is the propagator [Eq.~(\ref{propagator}) below], while the terms of order $\chi^n$ with $n\ge3$ in Eq.~(\ref{polymer field theory}) act as interaction vertices.
Throughout this paper, a ``vertex'' denotes the coefficient of such a higher-order term in the expansion in powers of $\chi$; in the diagrammatic representation used below, it is a square at which $n$ lines meet, and it carries the connected correlation function $MG_{\rm free}^{(n)}$ as shown in Fig.~\ref{fig:vertex}.
Note that (i) the $2$-point function (or propagator) is a function of the relative coordinate because of the translation symmetry and (ii) we can set the zero mode of $\chi(x)$ to zero since it only provides an overall factor in $\langle e^{-i\int d^Dx\hat{\rho}_S^{}(x)\chi(x)}\rangle_{\rm free}^{}$.
Equation~(\ref{polymer field theory}) is the polymer field theory we utilize in the following discussion.
It is also straightforward to obtain similar field theories for more general polymer systems, such as heteropolymers~\cite{Lin2016, McCarty2019, Wessen2022}.

It would be helpful to emphasize several differences compared to ordinary field theory.
In quantum field theory, each vertex contains a coupling constant, and we often perform perturbative calculations by expanding physical observables as functions of these coupling constants.
In the present polymer field theory, on the other hand, all microscopic coupling constants are contained in the polymer potential~(\ref{polymer potential}), which appears as the kinetic term in Eq.~(\ref{kinetic term}).
In other words, each vertex contains no free parameter except for the polymer density $\rho$ (and the bond length $b$) in the polymer field theory.
Physically, each vertex represents the connected $n$-point density correlation function of a free single polymer and thus encodes the conformational statistics originating from the chain connectivity, rather than an interaction.

Moreover, let us summarize relevant thermodynamic quantities~\cite{Hansen2013}.
The Helmholtz free energy is
\aln{
F=Vf=-\beta^{-1}\log Z~,
}
by which we can calculate the chemical potential as
\aln{
\mu=\frac{\partial F}{\partial M}=\frac{\partial f}{\partial \rho}~.
}
The pressure and isothermal compressibility are given by~\cite{Hansen2013, Fredrickson2023}
\aln{
p=-\frac{\partial F}{\partial V}=-f+\rho \mu~,\quad \kappa_T^{-1}&=\rho\frac{\partial p}{\partial \rho}=\rho^2\frac{\partial \mu}{\partial \rho}=\rho^2\frac{\partial^2 f}{\partial \rho^2}~.
}
The critical point $(\rho,T)=(\rho_c^{},T_c^{})$ is determined by
\aln{
\frac{\partial p}{\partial \rho}=\frac{\partial^2 p}{\partial \rho^2}=0\quad \leftrightarrow \quad \frac{\partial \mu}{\partial \rho}=\frac{\partial^2 \mu}{\partial \rho^2}=0~.
}
For $T<T_c^{}$, the system exhibits a first-order phase transition, which implies that there exist two (or multiple) densities that yield the same pressure $p(\rho_l)=p(\rho_h)$ and chemical potential $\mu(\rho_l)=\mu(\rho_h)$.
Hence, a binodal curve is determined by the simultaneous solution of
\aln{
p(\rho_l)=p(\rho_h)~,\quad \mu(\rho_l)=\mu(\rho_h)~
}
for a given temperature $T$~\cite{Landau1980, Delaney2017, Lin2023}.

\subsection{Random phase approximation}
\label{subsec_rpa}

The RPA corresponds to a $1$-loop approximation in field theory language.
More explicitly, the RPA free energy is obtained from Eq.~(\ref{polymer field theory}) by keeping only the Gaussian (quadratic) part of the action of $\chi(x)$ and performing the resulting Gaussian path integral, which yields the $\mathrm{Tr}\log$ contribution below~\cite{Fredrickson2006, Peskin1995}.
Namely, the $1$-loop free energy is
\aln{
f_{\rm RPA}^{}&=f_0^{}+\frac{N^2\tilde{v}(0)}{2}\rho^2+\frac{k_BT}{2}\mathrm{Tr}\left(\log \Gamma^{(2)}\right)
\label{RPA Trlog}
\\
&=f_0^{}+\frac{N^2\tilde{v}(0)}{2}\rho^2+\frac{k_BT}{2}\int \frac{d^Dk}{(2\pi)^D}\log\left(1+\beta \rho \tilde{v}(k)\tilde{g}_{\rm free}^{(2)}(k)\right)~,
\label{RPA free energy}
}
where $\tilde{v}(k)=\int d^Dxe^{-ik\cdot x}v(x)$ and we have omitted a constant term.
Equation~(\ref{RPA Trlog}) follows from the Gaussian path integral over the quadratic part of Eq.~(\ref{polymer field theory}), $\int{\cal D}\chi\,\exp\!\left(-\frac{1}{2}\int d^Dx\int d^Dy\,\chi(x)\Gamma^{(2)}(x-y)\chi(y)\right)\propto\left(\det\Gamma^{(2)}\right)^{-1/2}$, whose contribution to $f=-k_BT(\log Z)/V$ is $\frac{k_BT}{2V}\log\det\Gamma^{(2)}=\frac{k_BT}{2}\mathrm{Tr}\log\Gamma^{(2)}$, where $\mathrm{Tr}$ stands for $V^{-1}\int d^Dk/(2\pi)^D$ acting on the Fourier-diagonal kernel.
To pass to Eq.~(\ref{RPA free energy}), we use $\mathrm{Tr}\log\Gamma^{(2)}=\int d^Dk/(2\pi)^D\,\log\tilde{\Gamma}^{(2)}(k)$ with $\tilde{\Gamma}^{(2)}(k)=k_BT\tilde{v}(k)^{-1}+\rho\tilde{g}_{\rm free}^{(2)}(k)=k_BT\tilde{v}(k)^{-1}\left[1+\beta\rho\tilde{v}(k)\tilde{g}_{\rm free}^{(2)}(k)\right]$; the factor $k_BT\tilde{v}(k)^{-1}$ contributes a $\rho$-independent term, which is the constant omitted above.
By taking the $\rho$ derivative, the chemical potential is
\aln{
\mu_{\rm RPA}^{}=&k_BT\log\left(\frac{\rho}{(2\pi mk_BT/h^2)^{\frac{DN}{2}}(2\pi b^2/D)^{\frac{D(N-1)}{2}}}\right) - \frac{N}{2}v(0)
\nn
&+N^2\tilde{v}(0) \rho+\frac{1}{2}\int \frac{d^Dk}{(2\pi)^D}\frac{\tilde{v}(k)\tilde{g}_{\rm free}^{(2)}(k)}{1+\beta \rho \tilde{v}(k)\tilde{g}_{\rm free}^{(2)}(k)}~,
}
and the pressure is given by $p_{\rm RPA}^{}=-f_{\rm RPA}+\rho \mu_{\rm RPA}^{}$.

Here, we should comment on the convergence of these integrations.
For a short-range potential such as a Gaussian potential $\tilde{v}(k)\propto e^{-a^2k^2}$, these integrations are strongly convergent and there is no need for the renormalization.
On the other hand, for a long-range potential such as $\tilde{v}(k)\propto k^{-2}$, these integrations are divergent for $D\geq 2$, and such UV divergences must be subtracted (or renormalized) to obtain finite results.
Such a subtraction, however, does not change thermodynamic properties of the system.
We note that, in the analyses of Sec.~\ref{sec_comparison}, we do not perform any renormalization because we consider a superposition of short-range Gaussian potentials, for which the momentum integrations are convergent.
The present framework also applies to long-range interactions by adding the corresponding contribution to $\tilde{v}(k)$, with the only additional care being the UV subtraction discussed above; combining the short-range RPA+ with existing long-range RPA models for polyelectrolytes or charged proteins, however, requires extending the formulation to multicomponent or heteropolymer systems (see Ref.~\cite{Grzywacz2007} for renormalization for polymer mixtures).

\section{Beyond random phase approximation}
\label{sec_beyond}

\subsection{Loop expansion}
Here, we clarify that the expansion in powers of $\rho^{-1}$ corresponds to loop expansion, i.e., it corresponds to $\hbar$ in quantum field theory~\cite{Peskin1995}.
To this end, we express the effective Hamiltonian $S[\chi]$ for $\chi$, i.e., the exponent of Eq.~(\ref{polymer field theory}) with an overall minus sign, by the Fourier modes as
\aln{
\label{Fourier mode expansion}
S[\chi]&=\frac{1}{2}\int \frac{d^Dk}{(2\pi)^D}\tilde{\chi}(k)^*
\tilde{\Gamma}^{(2)}(k)\tilde{\chi}(k)~
\nn
-\rho&\sum_{n=3}^{\infty}\frac{(-i)^n}{n!}\left(\prod_{i=1}^n\int\frac{d^Dk_i^{}}{(2\pi)^D}\tilde{\chi}(k_i^{})\right)\tilde{g}^{(n)}_{\rm free}(\{k_i^{}\}_{i=1}^n)(2\pi)^D\delta^{(D)}\left(\sum_{i=1}^{n}k_i^{}\right)~,
}
where we have used Eq.~(\ref{Fourier modes of connected}) to replace $MG_{\rm free}^{(n)}$ by $\rho\,\tilde{g}_{\rm free}^{(n)}$.
Thus, Eq.~(\ref{Fourier mode expansion}) is nothing but the exponent of Eq.~(\ref{polymer field theory}) rewritten in terms of the Fourier modes, with an overall minus sign so that the statistical weight of $\chi$ is $e^{-S[\chi]}$.
Here, $S[\chi]$ is the effective Hamiltonian for the density-fluctuation field $\chi$ about the homogeneous saddle point~\cite{Fredrickson2006}, obtained by expanding $-M\log Q[-i\phi]$ (together with the Gaussian term) in powers of $\chi$: the quadratic part defines the propagator below, while each higher-order term of order $\chi^n$ involves the connected correlation function $\tilde{g}^{(n)}_{\rm free}$ and plays the role of an $n$-point vertex.
We keep the symbol $S[\chi]$ and the word ``action'' by analogy with the Euclidean action of quantum field theory, but in the language of polymer field theory, $S[\chi]$ is the effective Hamiltonian governing the density fluctuations.
Retaining only the quadratic part of Eq.~(\ref{Fourier mode expansion}), $\frac{1}{2}\int d^Dk/(2\pi)^D\,\tilde{\chi}(k)^*\tilde{\Gamma}^{(2)}(k)\tilde{\chi}(k)$, and using the standard Gaussian identity that the two-point average equals the inverse of the quadratic kernel, one obtains the propagator as
\aln{\label{propagator}
\langle\tilde{\chi}(k)^*\tilde{\chi}(q)\rangle=\frac{(2\pi)^D\delta^{(D)}(k-q)}{\tilde{\Gamma}^{(2)}(k)}=\frac{(2\pi)^D\delta^{(D)}(k-q)}{k_BT\tilde{v}(k)^{-1}+\rho\tilde{g}_{\rm free}^{(2)}(k)}~,
}
and see that each vertex is proportional to $\rho$.
Physically, the propagator $\langle\tilde{\chi}(k)^*\tilde{\chi}(q)\rangle$ is the Gaussian two-point average of the density-fluctuation field, i.e., the inverse of the quadratic operator $\tilde{\Gamma}^{(2)}(k)$, and corresponds to the density-density correlation function at the level of the RPA~\cite{Fredrickson2006}.

We now explain how the Feynman diagrams below encode the loop expansion of the free energy.
From Eq.~(\ref{field theory}) or (\ref{polymer field theory}), the free-energy density is given by $f=-k_BT(\log Z)/V$, and the only nontrivial part is the fluctuation path-integral $\log\int{\cal D}\chi\,e^{-S[\chi]}$.
We evaluate it perturbatively: the non-Gaussian ($n\ge3$) part of $S[\chi]$ is expanded out of the exponential, and the resulting polynomials in $\tilde{\chi}$ are averaged with the Gaussian weight $\exp[-\frac{1}{2}\int\frac{d^Dk}{(2\pi)^D}\tilde{\chi}(k)^*\tilde{\Gamma}^{(2)}(k)\tilde{\chi}(k)]$.
By Wick's theorem, each Gaussian average reduces to a sum of products of the two-point average [Eq.~(\ref{propagator})], and only the terms that remain connected contribute to $\log Z$~\cite{Peskin1995}.

A Feynman diagram is a compact notation for such a connected contribution, and its ingredients map directly onto the mathematics.
A line represents one propagator factor $\langle\tilde{\chi}^*\tilde{\chi}\rangle$ of Eq.~(\ref{propagator}).
A square vertex with $n$ legs represents the factor $\rho\,\tilde{g}_{\rm free}^{(n)}(k_1^{},\cdots,k_n^{})$, with the leg momenta as its arguments.
The momenta meeting at each vertex sum to zero (``momentum conservation,'' from the factor $(2\pi)^D\delta^{(D)}(\sum_i k_i^{})$ in each vertex), and every momentum not fixed by these constraints is integrated as $\int d^Dk/(2\pi)^D$, each such integration being one loop.
A rational prefactor (the symmetry factor) collects the $1/n!$ of each vertex and the number of equivalent Wick pairings.
%For example, the $(n,V_n^{})=(4,1)$ diagram in Fig.~\ref{fig:LO} consists of a single four-leg vertex $\rho\,\tilde{g}_{\rm free}^{(4)}$ whose legs are joined pairwise into two propagators; there are $3$ inequivalent pairings, so the prefactor is $3/4!=1/8$, and momentum conservation at the vertex leaves two loop momenta $k$ and $q$ to be integrated, giving $\Delta f^{(4,1)}$ in Sec.~\ref{subsec_rpa_plus}.

%Because we only need thermodynamic quantities, no separate tabulation of Feynman rules is required: at each order it is enough to list the diagrams and evaluate each by directly performing the Gaussian $\chi$ integrals in Eq.~(\ref{polymer field theory}).
The list of diagrams at each order is fixed by the power counting derived below: a connected vacuum (or bubble) diagram, i.e., one with no external legs, that contains $L$ independent loop integrations contributes as $\rho^{-(L-1)}$ [Eqs.~(\ref{V and P})--(\ref{rho dependence})].
Enumerating all corrections of order $\rho^{-(L-1)}$ therefore reduces to listing every connected vacuum diagram with $L$ loops that can be built from the vertices $\{\tilde{g}_{\rm free}^{(n)}\}_{n\ge3}$ and the propagators, equivalently every solution $\{(n,V_n^{})\}$ of the partition equation~(\ref{partition}) discussed below.
%Since this enumeration is finite, the set of diagrams obtained at each order is both correct and complete.

Now let us examine how each Feynman diagram contributes to the free energy.
Let us focus on the $n$th-order vertex $\rho\tilde{g}_{\rm free}^{(n)}$ and denote the number of vertices (propagators) by $V_n^{}~(\mathrm{P})$.
Since each propagator is constructed from $2$ legs, we have a relation
\aln{\label{V and P}
nV_n^{}=2\mathrm{P}~.
}
Then, the $\rho$ dependence is given by
\aln{
\frac{\rho^{V_n^{}}}{(\tilde{\Gamma}^{(2)})^{\mathrm{P}}}\sim \rho^{V_n^{}-\mathrm{P}}=\rho^{-\left(\frac{n}{2}-1\right)V_n^{}}\quad \text{for }\rho\rightarrow \infty~,
\label{rho dependence}
}
which clearly indicates that higher-loop diagrams are suppressed by a power of $\rho^{-1}$.
In addition, one can also check that the exponent in Eq.~(\ref{rho dependence}) is the same as the number of loop integrations: let us denote the number of loop integrations by $L$.
Since there are effectively $(n-1)$ momentum integrations for each vertex and each propagator preserves the momentum [see Eq.~(\ref{propagator})], we have
\aln{
L=(n-1)V_n^{}-\mathrm{P}+1=1+\left(\frac{n}{2}-1\right)V_n^{}~,
}
where $+1$ in the first equality represents the total momentum conservation, and we have used Eq.~(\ref{V and P}).
Now Eq.~(\ref{rho dependence}) is written as $\rho^{-(L-1)}$ which clearly indicates that $\rho^{-1}$ corresponds to $\hbar$ in quantum field theory.
See also Ref.~\cite{Morse2006} for a related power-counting argument in the diagrammatic expansion of polymer fluids.
Note also that since our polymer field theory~(\ref{polymer field theory}) is nothing but a (real) scalar field theory, it is in principle possible to construct its Feynman rules.
However, such rules are not necessarily required as long as we
focus on the calculations of thermodynamic quantities such as the free energy: it is sufficient to identify the diagrams appearing at a given order through the power counting above, and to evaluate each diagram by directly performing the perturbative Gaussian path-integrals in Eq.~(\ref{polymer field theory}).

When a diagram contains multiple types of vertices, the right hand side of Eq.~(\ref{rho dependence}) is generalized to
\aln{\rho^{-\sum_{n=3}^{\infty}(\frac{n}{2}-1)V_n^{}}~.
}
In particular, the LO corrections correspond to
\aln{
(n,V_n^{})=(4,1)~,\quad (3,2)~,
}
and the NLO corrections are
\aln{
(n,V_n^{})=&(6,1)~,\quad (4,2)~,\quad (3,4)~,
\nn
&(3,2)+(4,1)~,\quad \text{and}\quad (3,1)+(5,1)~.
}
In general, the problem is equivalent to counting the number of partitions of $2(L-1)$ in the form
\aln{
\sum_{n=3}^{\infty}(n-2)V_n^{}=2(L-1)~.
\label{partition}
}

Below, we explicitly calculate the LO and NLO corrections.

\subsection{Leading order corrections (RPA+)}
\label{subsec_rpa_plus}

We examine the LO corrections.
We refer to the extension of RPA that includes the LO corrections as RPA+.

\begin{figure}
\begin{center}
\includegraphics[scale=0.6]{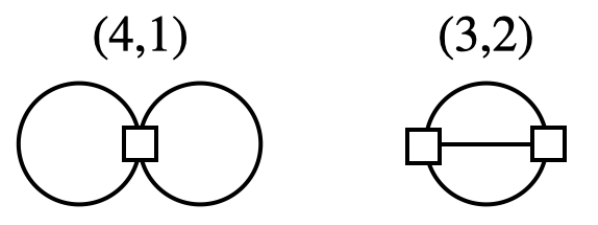}
\caption{
The LO diagrams contributing to the free energy.
Here, black lines correspond to the propagator (\ref{propagator}), and squares correspond to the connected correlation functions as shown in Fig.~\ref{fig:vertex}.
The set of integers above each diagram indicates $(n, V_n)$.}
\label{fig:LO}
\end{center}
\end{figure}

The Feynman diagram of the $(n,V)=(4,1)$ correction corresponds to the left one in Fig.~\ref{fig:LO}.
Evaluating this diagram as described above, it is given by
\aln{
&\Delta f^{(4,1)}=-\frac{k_BT\rho}{8}\int\frac{d^Dk}{(2\pi)^{D}}\int\frac{d^Dq}{(2\pi)^{D}}
\frac{\tilde{g}_{\rm free}^{(4)}(k^{},-k^{},q,-q)}{\tilde{\Gamma}^{(2)}(k)\tilde{\Gamma}^{(2)}(q)}
\\
&=-\frac{k_BT\rho}{8}\int\frac{d^Dk}{(2\pi)^{D}}\int\frac{d^Dq}{(2\pi)^{D}}
\frac{\tilde{f}_{\rm free}^{(4)}(k^{},-k^{},q,-q)-\tilde{f}_{\rm free}^{(2)}(k)\tilde{f}_{\rm free}^{(2)}(q)}{\tilde{\Gamma}^{(2)}(k)\tilde{\Gamma}^{(2)}(q)}~,
}
where we have used the relation between $G_{\rm free}^{(4)}$ and $F_{\rm free}^{(4)}$~(\ref{4-point relation}).

Next, the Feynman diagram of the $(3,2)$ correction corresponds to the right one in Fig.~\ref{fig:LO}.
It contributes to the free-energy density as
\aln{
\Delta f^{(3,2)}=k_BT\frac{\rho^2}{12}&\int\frac{d^Dk}{(2\pi)^D}\int\frac{d^Dq}{(2\pi)^D}
\nn
&\times \frac{\tilde{f}_{\rm free}^{(3)}(k,q,-k-q)\tilde{f}_{\rm free}^{(3)}(-k,-q,k+q)}{\tilde{\Gamma}^{(2)}(k)\tilde{\Gamma}^{(2)}(q)\tilde{\Gamma}^{(2)}(-k-q)}~,
}
where we have used $\tilde{f}_{\rm free}^{(3)}=\tilde{g}_{\rm free}^{(3)}$.
The complete LO correction to the free-energy density is the sum of these two diagrams, so that the RPA+ free-energy density reads
\aln{
f_{\rm RPA+}^{}=f_{\rm RPA}^{}+\Delta f^{(4,1)}+\Delta f^{(3,2)}~,
\label{rpa plus}
}
with $f_{\rm RPA}^{}$ given by Eq.~(\ref{RPA free energy}).
% analogously, $f_{\rm RPA++}^{}$ is obtained by further adding the NLO diagrams summarized in Appendix~\ref{app:NLO}.
The corrections to the chemical potential can be obtained by taking the $\rho$ derivative on these terms.
Note that performing these $2$-loop integrations analytically seems quite hard because each correlation function~(\ref{Fourier mode correlations}) contains many Gaussian terms as functions of wave vectors.
In this paper, we perform these integrations numerically (see Appendix~\ref{app_rpa_plus} for details).

\subsection{Next-to-leading order corrections (RPA++)}
\begin{figure}
\begin{center}
\includegraphics[scale=0.5]{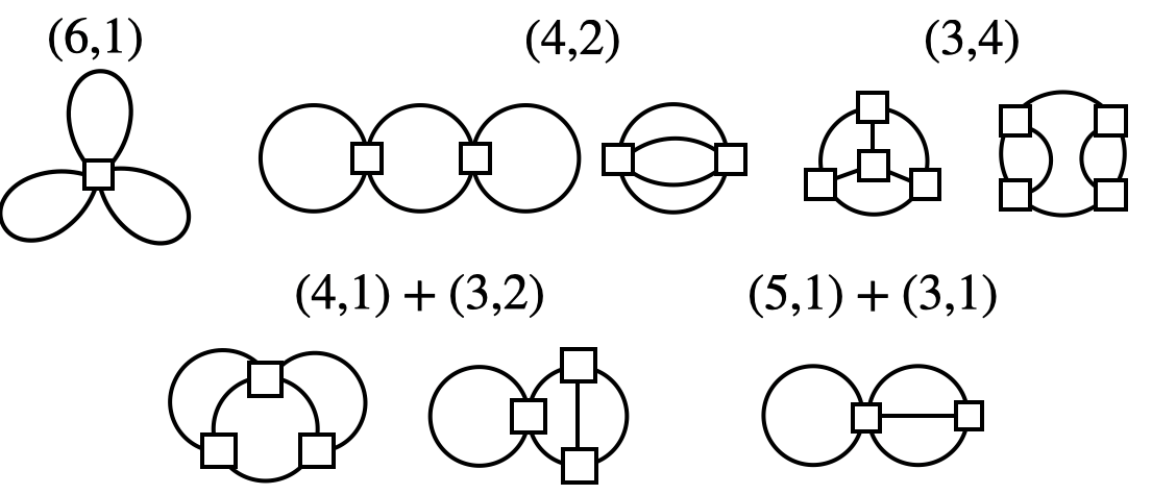}
\caption{The NLO diagrams contributing to the free energy.
The set of integers above each diagram indicates $(n, V_n)$.}
\label{fig:NLO}
\end{center}
\end{figure}

At NLO, there are $8$ bubble diagrams in terms of the connected vertices, as shown in Fig.~\ref{fig:NLO}.
The upper-left diagram corresponds to $(n,V_n^{})=(6,1)$, whereas the upper-second and -third diagrams correspond to $(4,2)$.
The upper-fourth and fifth diagram corresponds to $(3,4)$.
On the other hand, the lower-left two diagrams correspond to $(4,1)+(3,2)$, and the lower-right diagram corresponds to $(5,1)+(3,1)$.
Note that similar diagrams have been derived in a different context of field theory~\cite{Gaite2026}
The calculations of these diagrams are completely parallel to the LO calculations, and the results are summarized in Appendix~\ref{app:NLO}.
We refer to the approximation that incorporates terms up to the NLO corrections as RPA++.

Note that the numerical evaluation of the NLO diagrams is much more demanding than that of the LO diagrams: each diagram involves a three-fold momentum integral, i.e., a six-dimensional integral after reducing the angular part by the rotational symmetry (Appendix~\ref{app:NLO}), and its integrand contains $\tilde{f}_{\rm free}^{(n)}$ with $n$ up to $6$, each being a sum of ${\cal O}(N^n)$ Gaussian terms.
For this reason, we restrict the numerical analysis in Sec.~\ref{sec_comparison} to the RPA+ and leave an explicit numerical evaluation of the RPA++ as important future work.

\begin{figure}[t]
    \centering
    \includegraphics[scale=1]{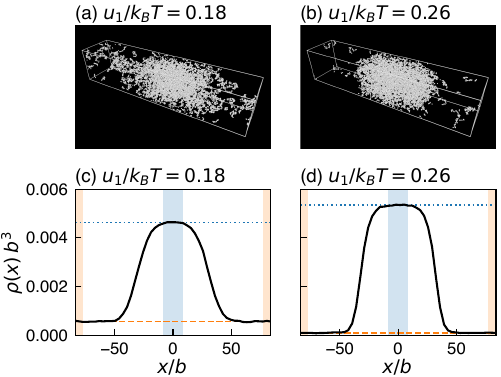}
    \caption{Simulation snapshots and density profile in a polymer solution model for $N = 30$ and $u_2 / u_1 = 0.1$.
    For (a), (c) $u_1 / k_B T = 0.18$ and (b), (d) $u_1 / k_B T = 0.26$, we show typical equilibrium snapshots (top panels) and the polymer density profile $\rho (x)$ (bottom panels).
    In panels (c) and (d), the central blue region and the outermost orange region suggest the domains to calculate coexistence densities $\rho_h$ (blue dotted line) and $\rho_l$ (orange dashed line), respectively.
    OVITO~\cite{Stukowski2010} was used for visualization in panels (a) and (b).}
    \label{fig_md_snapshots}
\end{figure}

\section{Comparison with molecular dynamics simulations}
\label{sec_comparison}

\subsection{Model and simulation protocol}

In the following, we consider three dimensional systems ($D=3$).
To test the predictions of the RPA and RPA+ against simulations, we performed MD simulations of a simple homopolymer solution model with implicit solvents.
Each polymer consists of $N = 30$ monomers connected by harmonic bonds with spring constant $3 k_B T / b^2$.
Non-bonded interactions between monomers are given by a superposition of Gaussian pair potentials of short-range repulsion and attraction:
\begin{equation}
    v(r) = u_1 e^{-r^2 / (4 {a_1}^2)} - u_2 e^{-r^2 / (4 {a_2}^2)},
    \label{eq_gaussian_model}
\end{equation}
where $a_i > 0$ and $u_i > 0$ ($i = 1, 2$) represent the interaction range and strength, respectively.
In the following, we use reduced units ($b = k_B T = m = 1$, where $m$ is the mass of a monomer) and fix $a_1 = 1$, $a_2 = 2$, and $u_2 / u_1 = 0.1$.
We vary $u_1$ as the effective inverse temperature of the system to examine the phase diagram in the $\rho$-$u_1$ plane.
Note that, with these parameter sets, the Fourier transformation $\tilde{v}(k)$ is positive for any $k$, particularly $\tilde{v}(0) > 0$, which satisfies a condition for thermodynamic stability~\cite{Ruelle1999, Fredrickson2023}.

To sample equilibrium configurations, we used Langevin dynamics simulations.
We set friction constant $\gamma = 1$ and timestep $\Delta t = 0.01$.
All the simulations were performed with HOOMD-blue~\cite{Anderson2020}.
As often used for studying phase separation in polymer models~\cite{Blas2008, Dignon2018, Tesei2021, Joseph2021, Feito2025}, we applied a slab geometry with a simulation box of aspect ratio $L_x : L_y : L_z = 5 : 1: 1$ under periodic boundary conditions, containing $M = 400$ or $800$ polymers.
We set the polymer density as $\rho b^3 = 2 \times 10^{-3}$.
For each value of $u_1 / k_B T \in \{ 0.16, 0.18, \ldots, 0.28, 0.30 \}$, we first relaxed the system through a short volume ramp and then recorded $200$ snapshots, each separated by $10^6$ steps (i.e., over $2 \times 10^8$ steps in total).
We discarded the first $20$ snapshots for equilibration and used the remaining $180$ as equilibrium configurations.

\begin{figure}[t]
    \centering
    \includegraphics[scale=1]{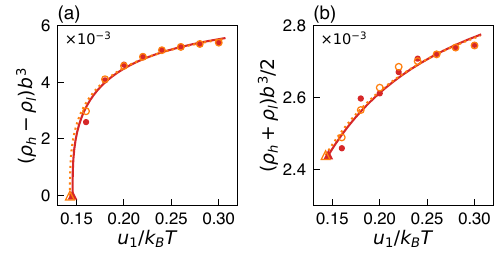}
    \caption{Fits used to estimate the critical point for $N = 30$ and $u_2 / u_1 = 0.1$: (a) the density difference $\rho_h - \rho_l$ and (b) the mean density $(\rho_h + \rho_l) / 2$ as functions of $u_1$, for $M = 400$ (red filled circles) and $M = 800$ (orange open circles).
    The red solid and orange dotted lines represent the fitted curves of Eqs.~(\ref{eq_ising_fit}) and (\ref{eq_diameter}) for $M = 400$ and $800$, respectively, and the triangles (red filled for $M = 400$ and orange open for $M = 800$) indicate the obtained critical point.}
    \label{fig_fit}
\end{figure}

\subsection{Extraction of coexistence densities}

For each snapshot, we divided the elongated $x$ axis into $50$ bins and counted the monomers in each bin to obtain the density profile $\rho(x)$.
The dense-phase density $\rho_h$ was estimated from the average over the central (i.e., around center-of-mass) $6$ bins containing the condensed slab, and the dilute-phase density $\rho_l$ was obtained by averaging over the outermost (i.e., farthest from the center-of-mass) $6$ bins.
We further took time average over the $180$ different configurations.

In Figs.~\ref{fig_md_snapshots}(a) and \ref{fig_md_snapshots}(b), we show typical snapshots of polymers in equilibrium at $u_1 = 0.18$ and $0.26$, respectively, where phase separation was observed.
The bottom panels [Figs.~\ref{fig_md_snapshots}(c) and \ref{fig_md_snapshots}(d)] represent the time-averaged density profile obtained for the corresponding parameters.
The dense and dilute domains used to calculate $\rho_h$ (blue dotted line) and $\rho_l$ (orange dashed line) are indicated as blue and orange regions, respectively.

\begin{figure}[t]
    \centering
    \includegraphics[scale=1]{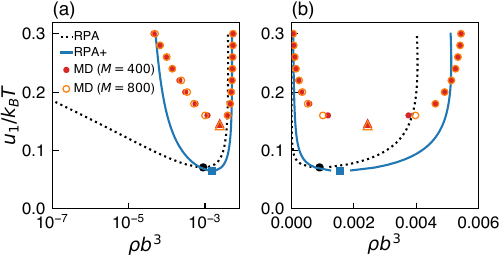}
    \caption{Binodal lines for phase separation in a polymer solution model for $N = 30$ and $u_2 / u_1 = 0.1$, plotted in the $\rho$-$u_1$ plane.
    Polymer density $\rho$ is shown in (a) logarithmic and (b) linear scales.
    The binodals predicted by the RPA (black dotted line) and the RPA+ (blue solid line) are compared with coexistence densities obtained from MD simulations with $M = 400$ polymers (red filled circles) and with a twice larger system, $M = 800$ (orange open circles).
    The critical points obtained from the RPA (black circle), RPA+ (blue square), and simulations (red filled triangle for $M = 400$ and orange open triangle for $M = 800$) are also plotted.}
    \label{fig_phase_diagram}
\end{figure}

The critical point $(\rho_c, u_{1c})$ was estimated by fitting the $u_1$ dependence of the coexistence densities to the Ising form~\cite{Feito2025},
\begin{equation}
    (\rho_h - \rho_l)^{1 / \beta_\mathrm{Ising}} = s_1 \bigg( 1 - \frac{u_{1c}}{u_1} \bigg),
    \label{eq_ising_fit}
\end{equation}
with $ \beta_\mathrm{Ising} = 0.326$, together with the rectilinear diameter rule
\begin{equation}
    \frac{\rho_h + \rho_l}{2} = \rho_c + s_2 \bigg( \frac{1}{u_1} - \frac{1}{u_{1c}} \bigg).
    \label{eq_diameter}
\end{equation}
Here, $s_1$ and $s_2$, as well as $\rho_c$ and $u_{1c}$, are fitting parameters.
The fitting results for $M = 400$ and $M = 800$ are shown in Fig.~\ref{fig_fit}.
Following this procedure, we obtained $\rho_c \approx 2.4 \times 10^{-3}$ and $u_{1c} \approx 0.15$ (see the triangles in Figs.~\ref{fig_fit} and \ref{fig_phase_diagram}).

\subsection{Comparison of binodal curves}

Figure~\ref{fig_phase_diagram} summarizes the comparison between the theoretical and simulated binodals.
As seen from the logarithmic-scale plot [Fig.~\ref{fig_phase_diagram}(a)], the binodal predicted by the RPA (black dotted line, Sec.~\ref{subsec_rpa})  extends into the low-density region $\rho < 10^{-5}$ and differs by several orders of magnitude from the dilute-phase density observed in MD simulations (red circles).
In contrast, the prediction of the binodal based on the RPA+ (blue solid line, Sec.~\ref{subsec_rpa_plus}) is closer to the observation, lying within the same order of magnitude.
This indicates that the lowest-order correction to the RPA qualitatively improves the prediction of thermodynamic properties.
In addition, as seen in the linear-scale plot [Fig.~\ref{fig_phase_diagram}(b)], the RPA+ improves the prediction of the dense-phase (high-density) coexistence density over the RPA, bringing it close to the values observed in simulations.
From Fig.~\ref{fig_phase_diagram}, we can confirm that finite-size effects are small compared to the deviation between the MD results and the theoretical predictions.

As seen from the linear-scale plot [Fig.~\ref{fig_phase_diagram}(b)], prediction of the binodal near the critical point is not improved significantly by the RPA+, compared to the RPA.
This suggests that higher-order corrections are increasingly important near the critical point, and we need other techniques such as the renormalization group to incorporate critical fluctuations and precisely estimate the critical point.

\subsection{Generality of improvements by RPA+}\label{subsec_generality}

\begin{figure}[t]
    \centering
    \includegraphics[scale=1]{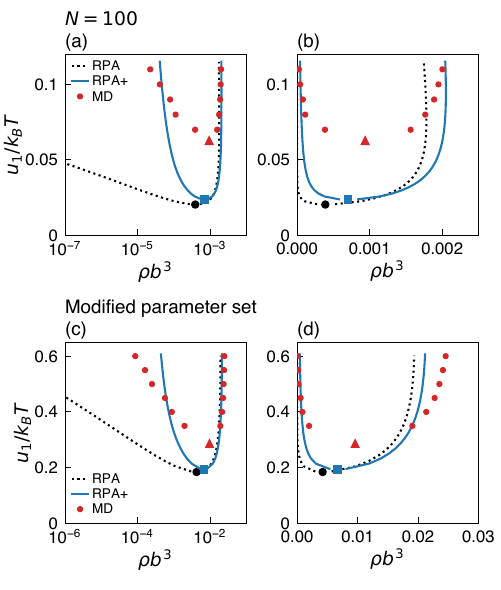}
    \caption{Binodal lines for [(a) and (b)] $N = 100$ with $u_2 / u_1 = 0.1$, $a_1 = 1$, and $a_2 = 2$ (i.e., the interaction parameters used for Fig.~\ref{fig_phase_diagram}) and [(c) and (d)] $N = 30$ with a modified parameter set, $u_2 / u_1 = 0.2$, $a_1 = 0.6$, and $a_2 = 1$, plotted in the $\rho$-$u_1$ plane.
    The polymer density $\rho$ is shown in [(a) and (c)] logarithmic and [(b) and (d)] linear scales.
    The binodals predicted by the RPA (black dotted line) and the RPA+ (blue solid line) are compared with coexistence densities obtained from MD simulations (red circles).
    The critical points obtained from the RPA (black circle), RPA+ (blue square), and simulations (red triangle) are also plotted.}
    \label{fig_generality}
\end{figure}

To examine the generality of the improvements by the RPA+, we performed additional MD simulations for (i) a longer chain length and (ii) a qualitatively different set of interaction parameters, using the same simulation protocol and analysis as above.

First, we increased the chain length to $N = 100$, keeping the interaction parameters unchanged.
We simulated $M = 100$ polymers at $\rho b^3 = 10^{-3}$ for $u_1 \in \{ 0.07, 0.08, \ldots, 0.11 \}$ and sampled $400$ snapshots, each separated by $10^6$ steps (i.e., over $4 \times 10^8$ steps in total).
We discarded the first $40$ snapshots for equilibration and used the remaining $360$ as equilibrium configurations.
As shown in Figs.~\ref{fig_generality}(a) and \ref{fig_generality}(b), the prediction of the RPA+ lies closer to the MD results than that of the RPA, whereas the underestimation of the critical point still persists.

Second, going back to $N = 30$, we employed a modified parameter set, $u_2 / u_1 = 0.2$, $a_1 = 0.6$, and $a_2 = 1$.
We simulated $M = 400$ polymers at $\rho b^3 = 8 \times 10^{-3}$ for $u_1 \in \{ 0.35, 0.40, \ldots, 0.60 \}$, sampling $200$ snapshots, each separated by $10^6$ steps (i.e., over $2 \times 10^8$ steps in total).
We discarded the first $20$ snapshots for equilibration and used the remaining $180$ as equilibrium configurations.
As shown in Figs.~\ref{fig_generality}(c) and \ref{fig_generality}(d), the RPA+ improves the binodal prediction over the RPA, whereas the underestimation of the critical point persists in this case as well.

These results indicate that the improvement of the binodal prediction by the RPA+ over the RPA is not specific to the parameter set examined in Fig.~\ref{fig_phase_diagram} but holds for broader sets of parameters.

\section{Conclusion and discussion}
\label{sec_conclusion}

In this paper, we developed a field-theoretic loop expansion of the free energy for polymer systems and clarified that it amounts to a systematic expansion in powers of $\rho^{-1}$, with the inverse polymer density $\rho^{-1}$ playing the role of the Planck constant $\hbar$ in quantum field theory.
We identified the leading-order (RPA+) and next-to-leading-order (RPA++) corrections and evaluated the RPA+ binodal for homopolymer solutions with short-range Gaussian interactions.
While the RPA++ corrections are derived (Appendix~\ref{app:NLO}), their explicit numerical evaluation is computationally demanding, and the quantitative analysis in this paper is limited to the RPA+; a numerical study of the RPA++ is left as important future work.
Compared with MD simulations, the RPA+ qualitatively improved the dilute-phase density predicted from the RPA by bringing it to the same order of magnitude as observed in simulations.
On the other hand, the critical point was not substantially improved, suggesting the need for more sophisticated methods to account for the critical fluctuations.

A further comment is in order before concluding this paper.
For a purely repulsive case, i.e., $u_2 = 0$ in Eq.~\eqref{eq_gaussian_model}, the homopolymer solution is well-mixed and should not exhibit phase separation, but the RPA predicts a spurious binodal with a critical point ($u_{1c} \approx 0.09$ for $N = 30$).
The leading-order correction (RPA+) does not cure this issue and still predicts a binodal line.
This indicates that, while systematic, the $\rho^{-1}$ expansion is not always qualitatively correct over the interaction parameter range.
It is therefore an important and interesting challenge to devise a method for systematically improving the $\rho^{-1}$ expansion so as to recover qualitatively correct predictions in such polymer systems.

Extending the present framework to heteropolymer phase separation relevant to biomolecular condensates~\cite{Lin2016, McCarty2019, Wessen2022, Lin2025} is also an interesting direction for future work.
This generalization should be obtained by making several quantities sequence-dependent: the two-body interaction becomes species-dependent, $v(x)\to v_{ab}(x)$ with $a,b$ labeling monomer species, so that $\tilde{v}(k)$ becomes a matrix $\tilde{v}_{ab}(k)$ carrying the sequence information; the free single-chain correlation functions are replaced by their sequence-resolved counterparts; and the propagator and vertices accordingly become species-dependent, so that the loop integrals involve matrix products and traces over species indices.
Nevertheless, the $\rho^{-1}$ power counting should remain unchanged.

Finally, we comment on the relation to other diagrammatic and perturbative expansions used in polymer physics.
Perturbation-theory and renormalization-group treatments of the excluded-volume interaction in a single chain can be cast into Feynman diagrams with excluded-volume vertices~\cite{Kholodenko1983} and have recently been extended to heteropolymers~\cite{Song2021}.
On the other hand, the lattice cluster theory~\cite{Nemirovsky1987, Dudowicz1991, Foreman1998} treats multiple-chain lattice systems by systematically expanding the free energy around the Flory-Huggins mean-field theory in powers of the inverse lattice coordination number and the reduced interaction energies, with the correction diagrams organized by the number of bonds.
In contrast to these expansions, our loop expansion is organized in powers of the inverse polymer density $\rho^{-1}$, which is a thermodynamically controllable parameter in multiple-chain systems, similarly to the method in Ref.~\cite{Morse2006}.
Which expansion is most adequate ultimately depends on the quantity of interest in multiple-polymer systems.

\section*{Acknowledgments}
We would like to thank K. Kawaguchi for fruitful discussions.
We thank J. Gaite for pointing out an overlooked NLO diagram in an earlier version of this paper.
This work was supported by JSPS KAKENHI Grant Numbers JP26K00052 and JP26K17111 (to K.A.).
Claude Opus and Fable were used to improve coding and readability.
This work is supported by KIAS Individual Grants, Grant No. 090901.

\appendix

\section{Monomer correlation functions}\label{app:monomer correlation}
The exponent in Eq.~(\ref{n-point Fourier mode}) can be written as
\aln{\sum_{j=1}^n k_j^{}\cdot x_{i_j^{}}^{}&=\sum_{j=1}^n k_j^{}\cdot \left(\sum_{l=1}^{i_j^{}-1}r_l^{}+x_1^{}
\right)=\sum_{j=1}^{n}k_j^{}\cdot x_1^{}+\sum_{j=1}^{n}\sum_{l=1}^{i_j^{}-1}k_j^{}\cdot r_l^{}
\nn
&=\sum_{j=1}^{n}k_j^{}\cdot x_1^{}+\sum_{j=1}^{n}\sum_{l=1}^{N-1}k_j^{}\cdot r_l^{}\theta(i_j^{}-1-l)
\nn
&=\sum_{j=1}^{n}k_j^{}\cdot x_1^{}+\sum_{l=1}^{N-1}K_{l}^{}\cdot r_l^{}
}
where $\theta(x)$ is the step function with $\theta(x)=1$ for $x\geq 0$ and $\theta(x)=0$ for $x<0$, $x_1^{}$ denotes the position of the first monomer that serves as a common reference for all the legs (upon integration, the term $\sum_{j}k_j^{}\cdot x_1^{}$ produces the momentum-conservation delta function), and
\aln{K_{l}^{}(\{i_j^{}\})\coloneqq \sum_{j=1}^{n} k_j^{}\theta(i_j^{}-1-l)~.
}
Here, the average in the single-polymer ensemble is taken over the position of the first monomer $x_1^{}$ and the bond vectors $\{r_l^{}\}_{l=1}^{N-1}$ with the Gaussian weight, namely
\aln{
\langle \cdots \rangle_{\rm free}^{}=\frac{1}{V}\int d^Dx_1^{}\,\frac{\left(\prod_{l=1}^{N-1}\int d^Dr_l^{}\right)e^{-\frac{D}{2b^2}\sum_{l=1}^{N-1}r_l^2}\,(\cdots)}{\left(\prod_{l=1}^{N-1}\int d^Dr_l^{}\right)e^{-\frac{D}{2b^2}\sum_{l=1}^{N-1}r_l^2}}~.
}
Thus, each $r_l^{}$ integration in the single polymer ensemble leads to
\aln{\frac{\int d^Dr_l^{}e^{-\frac{D}{2b^2}r_l^2-iK_{l}^{}(\{i_j^{}\})\cdot r_l^{}}}{\int d^Dr_l^{}e^{-\frac{D}{2b^2}r_l^2}}=e^{-\frac{b^2}{2D}K_l^{}(\{i_j^{}\})^2}~
}
using the well-known formula for the Gaussian integral.
Namely, we obtain
\aln{
\tilde{F}^{(n)}_{\rm free}(k_1^{},\cdots,k_n^{})=&\frac{1}{V}(2\pi)^D\delta^{(D)}\left(\sum_{j=1}^{n}k_j^{}\right)
\nn
\times\sum_{i_1^{}=1}^N\cdots \sum_{i_n^{}=1}^{N}&\exp\left(-\frac{b^2}{2D}\sum_{l=1}^{N-1}K_l^{}(\{i_j^{}\})^2
\right)~,
}
where $(2\pi)^D\delta^{(D)}\left(\sum_{j=1}^{n}k_j^{}\right)$ comes from the translation invariance.
By definition,
\aln{\sum_{l=1}^{N-1}K_l^{}(\{i_j^{}\})^2&=\sum_{l=1}^{N-1}\sum_{j=1}^{n} \sum_{j'=1}^{n} k_j^{}\cdot k_{j'}^{}\theta(i_j^{}-1-l)\theta(i_{j'}^{}-1-l)
\nn
&=\sum_{j=1}^{n} \sum_{j'=1}^{n}k_j^{}\cdot k_{j'}^{}\mathrm{min}(i_j^{}-1,i_{j'}^{}-1)~,}
which leads to Eq.~(\ref{Fourier mode correlations}).

\section{Relations among correlation functions}\label{app:correlation}
In this Appendix, we summarize the relations between $F_{\rm free}^{(n)}$ and $G_{\rm free}^{(n)}$ up to $n=4$.
See also Ref.~\cite{Peskin1995} for the detailed derivations.
For $n=1$,
\aln{
G_{\rm free}^{(1)}=F_{\rm free}^{(1)}=\frac{N}{V}~,
}
which is nothing but the monomer density in a single polymer.
For $n=2$,
\aln{F_{\rm free}^{(2)}(x_1^{},x_2^{})=G_{\rm free}^{(2)}(x_1^{},x_2^{})+(G_{\rm free}^{(1)})^2~,
}
and their Fourier modes satisfy
\aln{
\tilde{F}_{\rm free}^{(2)}(k_1^{},k_2^{})=\tilde{G}_{\rm free}^{(2)}(k_1^{},k_2^{})+(G_{\rm free}^{(1)})^2(2\pi)^{2D}\delta^{(D)}(k_1^{})\delta^{(D)}(k_2^{})~.
}
For $n=3$, we have
\aln{
F_{\rm free}^{(3)}(x_1^{},x_2^{},x_3^{})=&G_{\rm free}^{(3)}(x_1^{},x_2^{},x_3^{})+G_{\rm free}^{(1)}\bigg(G_{\rm free}^{(2)}(x_1^{},x_2^{})
\nn
&+G_{\rm free}^{(2)}(x_2^{},x_3^{})+G_{\rm free}^{(2)}(x_1^{},x_3^{})\bigg)+(G_{\rm free}^{(1)})^3~,
}
and their Fourier modes satisfy
\aln{
\tilde{F}_{\rm free}^{(3)}(k_1^{},k_2^{},k_3^{})&=\tilde{G}_{\rm free}^{(3)}(k_1^{},k_2^{},k_3^{})
+(2\pi)^{D}G_{\rm free}^{(1)}\bigg[\delta^{(D)}(k_1^{})\tilde{G}_{\rm free}^{(2)}(k_2^{},k_3^{})
\nn
&+\delta^{(D)}(k_2^{})\tilde{G}_{\rm free}^{(2)}(k_3^{},k_1^{})+\delta^{(D)}(k_3^{})\tilde{G}_{\rm free}^{(2)}(k_1^{},k_2^{})\bigg]
\nn
&+(G_{\rm free}^{(1)})^3(2\pi)^{3D}\delta^{(D)}(k_1^{})\delta^{(D)}(k_2^{})\delta^{(D)}(k_3^{})~.
}
For $n=4$, we have
\aln{
&F_{\rm free}^{(4)}(x_1^{},\cdots,x_4^{})=G_{\rm free}^{(4)}(x_1^{},\cdots,x_4^{})
\nn
&+G_{\rm free}^{(1)}\sum_{i=1}^4G_{\rm free}^{(3)}(x_1^{},\cdots,\check{x}_i^{},\cdots,x_4^{})
\nn
&+\sum_{i\neq j\neq k\neq l}G_{\rm free}^{(2)}(x_i^{},x_j^{})G_{\rm free}^{(2)}(x_k^{},x_l^{})
\nn
&+(G_{\rm free}^{(1)})^2\sum_{i\neq j}G_{\rm free}^{(2)}(x_i^{},x_j^{})+(G_{\rm free}^{(1)})^4~,
\label{4-point relation}
}
where the summations are understood to run over distinct partitions of the external points, and the inverted hat in $G_{\rm free}^{(3)}(x_1^{},\cdots,\check{x}_i^{},\cdots,x_4^{})$ indicates that the argument $x_i^{}$ is omitted from the list, so that this $G_{\rm free}^{(3)}$ is a function of the three remaining coordinates.

In particular, in terms of the reduced Fourier modes $\tilde{f}_{\rm free}^{(n)}$ and $\tilde{g}_{\rm free}^{(n)}$ defined in Eqs.~(\ref{Fourier modes of nonconnected}) and (\ref{Fourier modes of connected}), the above relations allow us to express $\tilde{g}_{\rm free}^{(n)}$ by $\tilde{f}_{\rm free}^{(n)}$.
These relations hold for arbitrary wave vectors. In the following we do not write down the most general expression; instead, we record $\tilde{g}_{\rm free}^{(n)}$ only in the specific momentum configurations that enter the LO diagrams of Sec.~\ref{subsec_rpa_plus}, namely $(k,-k)$ for $n=2$, a generic triangle $(k_1^{},k_2^{},k_3^{})$ with $\sum_i k_i^{}=0$ for $n=3$, and $(k,-k,q,-q)$ for $n=4$.
For external momenta such that no partial sum vanishes (except for the total momentum conservation), the terms involving $\delta^{(D)}$ of individual momenta do not contribute (these $\delta^{(D)}$ terms are supported on lower-dimensional subsets of the loop-integration domain and therefore drop out of the momentum integrals), and we obtain
\aln{
\tilde{g}_{\rm free}^{(2)}(k,-k)&=\tilde{f}_{\rm free}^{(2)}(k,-k)~,
\\
\tilde{g}_{\rm free}^{(3)}(k_1^{},k_2^{},k_3^{})&=\tilde{f}_{\rm free}^{(3)}(k_1^{},k_2^{},k_3^{})~.
}
For $n=4$ with the external momenta $(k,-k,q,-q)$ and $k\neq \pm q$, only the pairing $G_{\rm free}^{(2)}G_{\rm free}^{(2)}$ whose momenta sum to zero within each pair survives in the thermodynamic limit, which leads to
\aln{
\tilde{g}_{\rm free}^{(4)}(k,-k,q,-q)=\tilde{f}_{\rm free}^{(4)}(k,-k,q,-q)-\tilde{f}_{\rm free}^{(2)}(k,-k)\tilde{f}_{\rm free}^{(2)}(q,-q)~,
}
as used in Sec.~\ref{subsec_rpa_plus} with the shorthand notation $\tilde{f}_{\rm free}^{(2)}(k)\equiv \tilde{f}_{\rm free}^{(2)}(k,-k)$.

In the same way, we can obtain similar relations for higher-order correlation functions.
In particular, one can check that the zero mode of $G_{\rm free}^{(n)}$ is zero for $^\forall n\geq 2$.

\begin{widetext}

\section{Numerical evaluation of the RPA+ corrections}\label{app_rpa_plus}

In this Appendix, we describe the numerical evaluation of the LO (RPA+) corrections $\Delta f^{(4,1)}$ and $\Delta f^{(3,2)}$ in Sec.~\ref{subsec_rpa_plus}, restricting to $D=3$ as used in Sec.~\ref{sec_comparison}.
We write the three-dimensional wave vectors as $\vec{k}$ and $\vec{q}$, and their magnitudes as $k=|\vec{k}|$ and $q=|\vec{q}|$.
First, the propagator factor $1/\tilde{\Gamma}^{(2)}(\vec{k})$ in Eq.~(\ref{propagator}) can be rewritten as
\aln{
\frac{1}{\tilde{\Gamma}^{(2)}(\vec{k})}=\beta\tilde{v}(k)P(k)~,\quad P(k)\coloneqq \frac{1}{1+\beta\rho\tilde{v}(k)\tilde{g}_{\rm free}^{(2)}(k)}~,
}
where the two-point function has the closed form
\aln{
\tilde{g}_{\rm free}^{(2)}(k)=\sum_{n_1^{},n_2^{}=0}^{N-1}\Phi^{|n_1^{}-n_2^{}|}=\frac{N(1-\Phi^2)-2\Phi(1-\Phi^N)}{(1-\Phi)^2}~,\quad \Phi\coloneqq e^{-\frac{b^2k^2}{6}}~,
\label{g2 closed}
}
as obtained from Eq.~(\ref{Fourier mode correlations}).
The closed form on the right-hand side of Eq.~(\ref{g2 closed}) follows by writing the double sum as $\sum_{n_1^{},n_2^{}=0}^{N-1}\Phi^{|n_1^{}-n_2^{}|}=N+2\sum_{m=1}^{N-1}(N-m)\Phi^m$ and evaluating the finite geometric sums $\sum_{m=1}^{N-1}\Phi^m=\Phi(1-\Phi^{N-1})/(1-\Phi)$ and $\sum_{m=1}^{N-1}m\Phi^m$ with the standard formulas.
Since the vertex factors in $\Delta f^{(4,1)}$ and $\Delta f^{(3,2)}$ depend only on the magnitudes $k$ and $q$ and the relative angle $\theta$ between $\vec{k}$ and $\vec{q}$, the rotational symmetry reduces the $2$-loop momentum integrations to three-dimensional ones:
\aln{
\int \frac{d^3\vec{k}}{(2\pi)^3}\int \frac{d^3\vec{q}}{(2\pi)^3}=\frac{1}{8\pi^4}\int_0^{\infty}dk\,k^2\int_0^{\infty}dq\,q^2\int_{-1}^{1}dc~,\quad c\coloneqq \cos\theta~.
}
The LO corrections in Sec.~\ref{subsec_rpa_plus} then become
\aln{
\Delta f^{(4,1)}&=-\frac{\rho\beta}{8}\frac{1}{8\pi^4}\int_0^{\infty}dk\,k^2\int_0^{\infty}dq\,q^2\int_{-1}^{1}dc~
\tilde{v}(k)\tilde{v}(q)P(k)P(q)\,S(k,q,c)~,
\label{app_df41}
\\
\Delta f^{(3,2)}&=\frac{\rho^2\beta^2}{12}\frac{1}{8\pi^4}\int_0^{\infty}dk\,k^2\int_0^{\infty}dq\,q^2\int_{-1}^{1}dc~
\tilde{v}(k)\tilde{v}(q)\tilde{v}(k_+^{})P(k)P(q)P(k_+^{})\,\tilde{S}(k,q,c)~,
\label{app_df32}
}
with $k_+^{}\coloneqq |\vec{k}+\vec{q}|=\sqrt{k^2+q^2+2kqc}$.
Here, the vertex factors $S(k,q,c)\coloneqq \tilde{g}_{\rm free}^{(4)}(\vec{k},-\vec{k},\vec{q},-\vec{q})$ and $\tilde{S}(k,q,c)\coloneqq \tilde{f}_{\rm free}^{(3)}(\vec{k},\vec{q},-\vec{k}-\vec{q})\tilde{f}_{\rm free}^{(3)}(-\vec{k},-\vec{q},\vec{k}+\vec{q})$ are obtained from Eq.~(\ref{Fourier mode correlations}) and the relations in Appendix~\ref{app:correlation} as
\aln{
S(k,q,c)&=\sum_{n_1^{},\cdots,n_4^{}=0}^{N-1}\left[e^{-\frac{b^2}{6}\left(k^2|n_1^{}-n_2^{}|+q^2|n_3^{}-n_4^{}|+2kqc\,m_4^{}\right)}-e^{-\frac{b^2}{6}\left(k^2|n_1^{}-n_2^{}|+q^2|n_3^{}-n_4^{}|\right)}\right]~,
\\
m_4^{}&\coloneqq \mathrm{min}(n_1^{},n_3^{})+\mathrm{min}(n_2^{},n_4^{})-\mathrm{min}(n_1^{},n_4^{})-\mathrm{min}(n_2^{},n_3^{})~,
\\
\tilde{S}(k,q,c)&=\left[\sum_{n_1^{},n_2^{},n_3^{}=0}^{N-1}e^{-\frac{b^2}{6}\left(k^2|n_1^{}-n_3^{}|+q^2|n_2^{}-n_3^{}|+2kqc\,m_3^{}\right)}\right]^2~,
\\
m_3^{}&\coloneqq n_3^{}+\mathrm{min}(n_1^{},n_2^{})-\mathrm{min}(n_1^{},n_3^{})-\mathrm{min}(n_2^{},n_3^{})~,
}
where we have used $\tilde{f}_{\rm free}^{(3)}(-\vec{k},-\vec{q},\vec{k}+\vec{q})=\tilde{f}_{\rm free}^{(3)}(\vec{k},\vec{q},-\vec{k}-\vec{q})$, which follows from Eq.~(\ref{Fourier mode correlations}).

Note that $S$ and $\tilde{S}$ encode only the conformational statistics of a free single chain and are independent of the polymer density $\rho$ and the interaction potential $\tilde{v}$.
We thus precompute $S$ and $\tilde{S}$ on a grid of $(k,q,c)$ once for a given $N$ and reuse them for all densities and interaction parameters.
For the results in Sec.~\ref{sec_comparison}, we discretized $k,q\in[10^{-4},8]$ (in units of $b^{-1}$) with $40$ uniform points and $c\in[-1,1]$ with $20$ uniform points, and evaluated the integrals~(\ref{app_df41}) and (\ref{app_df32}) with the trapezoidal rule.
Since $\rho$ enters Eqs.~(\ref{app_df41}) and (\ref{app_df32}) only through $P$ and the explicit prefactors, the corrections to the chemical potential $\Delta \mu=\partial \Delta f/\partial \rho$ and to the pressure $\Delta p=\rho\Delta \mu-\Delta f$, as well as the derivatives $\partial \Delta \mu/\partial \rho$ and $\partial^2 \Delta \mu/\partial \rho^2$ needed to locate the critical point, are obtained analytically by using $\partial P(k)/\partial \rho=-\beta\tilde{v}(k)\tilde{g}_{\rm free}^{(2)}(k)P(k)^2$ and evaluated on the same grid.
For each $u_1$, we located the binodal from the crossing of the parametric $(\mu ,p)$ curve when varying $\rho$, as used in previous studies~\cite{Delaney2017, Lin2023}.

\section{NLO corrections} \label{app:NLO}
In this Appendix, we summarize the NLO corrections.
First, the $(6,1)$ contribution is
\aln{
\Delta f^{(6,1)}/k_BT&=\frac{\rho}{48}
\int\frac{d^Dk}{(2\pi)^D}\int\frac{d^Dq}{(2\pi)^D}\int\frac{d^Dp}{(2\pi)^D} \frac{\tilde{g}_{\rm free}^{(6)}(k,-k,q,-q,p,-p)}{\tilde{\Gamma}^{(2)}(k)\tilde{\Gamma}^{(2)}(q)\tilde{\Gamma}^{(2)}(p)}
\\
&=\rho\bigg[\frac{1}{48}
\int\frac{d^Dk}{(2\pi)^D}\int\frac{d^Dq}{(2\pi)^D}\int\frac{d^Dp}{(2\pi)^D} \frac{\tilde{f}_{\rm free}^{(6)}(k,-k,q,-q,p,-p)}{\tilde{\Gamma}^{(2)}(k)\tilde{\Gamma}^{(2)}(q)\tilde{\Gamma}^{(2)}(p)}
\nn
 &\hspace{2cm}-\frac{1}{16}\left(\int\frac{d^Dk}{(2\pi)^D}\frac{\tilde{g}_{\rm free}^{(2)}(k)}{\tilde{\Gamma}^{(2)}(k)}
\right)\left(\int\frac{d^Dp}{(2\pi)^{D}}\int\frac{d^Dq}{(2\pi)^{D}}\frac{\tilde{f}_{\rm free}^{(4)}(p,-p,q,-q)}{\tilde{\Gamma}^{(2)}(p)\tilde{\Gamma}^{(2)}(q)}\right)
+\frac{1}{24}\left(\int\frac{d^Dk}{(2\pi)^D}\frac{\tilde{g}_{\rm free}^{(2)}(k)}{\tilde{\Gamma}^{(2)}(k)}
\right)^3
\bigg]~.
}
Next, the second and third diagrams in Fig.~\ref{fig:NLO} correspond to the $(4,2)$ contributions.
In terms of the non-connected correlation functions, they are
\aln{
\Delta f^{(4,2)}/k_BT=-\rho^2
\frac{1}{16}\int\frac{d^Dk}{(2\pi)^{D}}\int\frac{d^Dq}{(2\pi)^{D}}\int\frac{d^Dp}{(2\pi)^{D}}\bigg[&
\frac{\left(\tilde{f}_{\rm free}^{(4)}(k,-k,q,-q)-\tilde{f}_{\rm free}^{(2)}(k)\tilde{f}_{\rm free}^{(2)}(q)\right)\left(\tilde{f}_{\rm free}^{(4)}(q,-q,p,-p)-\tilde{f}_{\rm free}^{(2)}(q)\tilde{f}_{\rm free}^{(2)}(p)\right)}{\tilde{\Gamma}^{(2)}(k)\tilde{\Gamma}^{(2)}(q)^2\tilde{\Gamma}^{(2)}(p)}
\nn
&+\frac{1}{3}
\frac{(\tilde{f}_{\rm free}^{(4)}(k,q,p,-k-q-p))^2}{\tilde{\Gamma}^{(2)}(k)\tilde{\Gamma}^{(2)}(q)\tilde{\Gamma}^{(2)}(p)\tilde{\Gamma}^{(2)}(-k-q-p)}
\bigg]~.
}
The fourth and fifth diagram in Fig.~\ref{fig:NLO} corresponds to the $(3,4)$ contributions.
They are
\aln{
\Delta f^{(3,4)}/k_BT=-\rho^4\bigg[&\frac{1}{24}\int\frac{d^Dk}{(2\pi)^{D}}\int\frac{d^Dq}{(2\pi)^{D}}\int\frac{d^Dp}{(2\pi)^{D}}
\frac{
\tilde{f}_{\rm free}^{(3)}(k,q+p-k,-q-p)\tilde{f}_{\rm free}^{(3)}(q+p,-q,-p)
\tilde{f}_{\rm free}^{(3)}(p,q-k,k-q-p)
\tilde{f}_{\rm free}^{(3)}(q,-k,k-q)
}{\tilde{\Gamma}^{(2)}(k)\tilde{\Gamma}^{(2)}(q)\tilde{\Gamma}^{(2)}(p)\tilde{\Gamma}^{(2)}(q-k)\tilde{\Gamma}^{(2)}(q+p-k)\tilde{\Gamma}^{(2)}(q+p)}
\nn
&+\frac{1}{16}\int\frac{d^Dk}{(2\pi)^{D}}\int\frac{d^Dq}{(2\pi)^{D}}\int\frac{d^Dp}{(2\pi)^{D}}
\frac{
\tilde{f}_{\rm free}^{(3)}(k,q,-k-q)^2\tilde{f}_{\rm free}^{(3)}(k+q,p,-p-k-q)
}{\tilde{\Gamma}^{(2)}(k)\tilde{\Gamma}^{(2)}(q)\tilde{\Gamma}^{(2)}(k+q)^2\tilde{\Gamma}^{(2)}(p)\tilde{\Gamma}^{(2)}(q+p+k)}\bigg]~.
}
The sixth and seventh diagrams in Fig.~\ref{fig:NLO} correspond to the $(3,2)+(4,1)$ contributions.
Repeating a similar calculation, we obtain
\aln{
\Delta f^{(3,2)+(4,1)}/k_BT=\frac{\rho^3}{8}&\int\frac{d^Dk}{(2\pi)^{D}}\int\frac{d^Dq}{(2\pi)^{D}}\int\frac{d^Dp}{(2\pi)^{D}}
 \bigg(\frac{\tilde{f}_{\rm free}^{(4)}(k,q,p,-k-q-p)\tilde{f}_{\rm free}^{(3)}(-k,-q,k+q)\tilde{f}_{\rm free}^{(3)}(-p,-k-q,k+q+p)}{\tilde{\Gamma}^{(2)}(k)\tilde{\Gamma}^{(2)}(q)\tilde{\Gamma}^{(2)}(p)\tilde{\Gamma}^{(2)}(k+q)\tilde{\Gamma}^{(2)}(-k-q-p)}
\nn
+&
\frac{(\tilde{f}_{\rm free}^{(4)}(k,-k,q,-q)-\tilde{f}_{\rm free}^{(2)}(k)\tilde{f}_{\rm free}^{(2)}(q))\tilde{f}_{\rm free}^{(3)}(-q,p,q-p)\tilde{f}_{\rm free}^{(3)}(q,-p,-q+p)}{\tilde{\Gamma}^{(2)}(k)\tilde{\Gamma}^{(2)}(q)^2\tilde{\Gamma}^{(2)}(p)\tilde{\Gamma}^{(2)}(q-p)}
\bigg)~.
}
Finally, the last diagram in Fig.~\ref{fig:NLO} corresponds to the $(3,1)+(5,1)$ contribution.
It contributes to the free-energy density as
\aln{
\Delta f^{(3,1)+(5,1)}/k_BT=-\frac{\rho^2}{12}\bigg[&\int\frac{d^Dk}{(2\pi)^{D}}\int\frac{d^Dq}{(2\pi)^{D}}\int\frac{d^Dp}{(2\pi)^{D}}
 \frac{\tilde{f}_{\rm free}^{(5)}(k,-k,q,p,-q-p)\tilde{f}_{\rm free}^{(3)}(-q,-p,q+p)}{\tilde{\Gamma}^{(2)}(k)\tilde{\Gamma}^{(2)}(q)\tilde{\Gamma}^{(2)}(p)\tilde{\Gamma}^{(2)}(-q-p)}
\nn
&-\left(\int\frac{d^Dk}{(2\pi)^{D}}
\frac{\tilde{g}_{\rm free}^{(2)}(k)}{\tilde{\Gamma}^{(2)}(k)}\right)\left(
\int\frac{d^Dq}{(2\pi)^{D}}\int\frac{d^Dp}{(2\pi)^{D}}
\frac{(\tilde{f}_{\rm free}^{(3)}(q,p,-q-p))^2}{\tilde{\Gamma}^{(2)}(q)\tilde{\Gamma}^{(2)}(p)\tilde{\Gamma}^{(2)}(-q-p)}
\right)
\bigg]~.
}
By using the rotational symmetry, these momentum integrations can be reduced as follows for $D=3$.
We identify the direction of $\vec{k}$ as the $z$ axis and can take $\vec{q}$ to lie in the $x$-$z$ plane without loss of generality.
Then, these three momenta are explicitly parameterized as
\aln{
\vec{k}=k(0,0,1)~,\quad
\vec{q}=q(\sin\theta,0,\cos\theta)~,\quad \vec{p}=p(\sin\theta'\cos\varphi,\sin\theta'\sin\varphi,\cos\theta')~,
}
by which we can calculate their inner products as
\aln{
&\vec{k}\cdot \vec{q}=kq\cos\theta~,\quad
\vec{k}\cdot \vec{p}=kp\cos\theta'~,
\quad \vec{q}\cdot \vec{p}=qp(\cos\theta\cos\theta'+\sin\theta \sin\theta'\cos\varphi)~.
}
The corresponding integration measure is
\aln{
\int d^3\vec{k}\int d^3\vec{q}\int d^3\vec{p}=8\pi^2 \int_0^{\infty}dkk^2 \int_0^{\infty}dqq^2 \int_0^{\infty}dpp^2\int_{-1}^{1} d(\cos\theta)\int_{-1}^{1} d(\cos\theta')\int_{0}^{2\pi} d\varphi~.
}
Although these expressions are ready for numerical evaluation, the resulting six-dimensional integrations are computationally demanding because the integrands contain the correlation functions $\tilde{f}_{\rm free}^{(n)}$ with $n$ up to $6$, each being a sum of ${\cal O}(N^n)$ Gaussian terms (over $10^8$ terms already for $N=30$).
An explicit numerical evaluation of the RPA++ is left as future work.

\end{widetext}

\bibliography{Bibliography}

\end{document}